\PassOptionsToPackage{english}{babel}
\documentclass[useAMS,usenatbib]{mn2e}

\usepackage[english]{babel}
\usepackage{color}
\usepackage{graphicx}
\usepackage{subfigure}
\usepackage{adjustbox}
\usepackage{breqn}
\usepackage{tabu}
\usepackage[draft]{hyperref}
\title[Observability of CPDs]{Observability of Forming Planets and their Circumplanetary Disks I. -- Parameter Study for ALMA}
\author[Szul\'agyi et al.]{
\parbox{\textwidth}{
J. Szul\'agyi$^{1}$\thanks{E-mail:
judits@phys.ethz.ch}, G. van der Plas$^{2}$, M. R. Meyer$^{3}$, A. Pohl$^{4}$, S. P. Quanz$^{1}$,\\
L. Mayer$^{5}$, S. Daemgen$^{1}$, V. Tamburello$^{5}$
}\vspace{3mm}\\
$^{1}$ ETH Z\"urich, Institute for Particle Physics and Astrophysics, Wolfgang-Pauli-Strasse 27, CH-8093, Z\"urich, Switzerland\\
$^{2}$ Universit\'e Grenoble Alpes, IPAG, F-38000 Grenoble, France\\
$^{3}$ Department of Astronomy, University of Michigan, Ann Arbor, MI 48109, U.S.A.\\
$^{4}$ Max Planck Institute for Astronomy, K\"onigstuhl 17, 69117, Heidelberg, Germany \\
$^{5}$ Center for Theoretical Astrophysics and Cosmology, Institute for Computational Science, University of Zurich,\\
Winterthurerstrasse 190, CH-8057 Z\"urich, Switzerland\\
}

\begin{document}
 \pdfpagewidth 8.267in 
  \pdfpageheight 11.692in

\date{Accepted XX. Received XX; in original form 2017 February 3}

\pagerange{\pageref{firstpage}--\pageref{lastpage}} \pubyear{2017}

\maketitle

\label{firstpage}

\begin{abstract}
We present mock observations of forming planets with ALMA. The possible detections of circumplanetary disks (CPDs) were investigated around planets of Saturn, 1, 3, 5, and 10 Jupiter-masses that are placed at 5.2 AU from their star. The radiative, three dimensional hydrodynamic simulations were then post-processed with RADMC3D and the ALMA Observation Simulator. We found that even though the CPDs are too small to be resolved, they are hot due to the accreting planet in the optically thick limit, therefore the best chance to detect them with continuum observations in this case is at the shortest ALMA wavelengths, such as Band 9 (440 microns). Similar fluxes were found in the case of Saturn and Jupiter-mass planets, as for the 10 $\mathrm{M_{Jup}}$ gas-giant, due to temperature weighted optical depth effects: when no deep gap is carved, the planet region is blanketed by the optically thick circumstellar disk leading to a less efficient cooling there. A test was made for a 52 AU orbital separation, showed that optically thin CPDs are also detectable in band 7 but they need longer integration times ($>$5hrs). Comparing the gap profiles of the same simulation at various ALMA bands and the hydro simulation confirmed that they change significantly, first because the gap is wider at longer wavelengths due to decreasing optical depth; second, the beam convolution makes the gap shallower and at least 25\% narrower. Therefore, caution has to be made when estimating planet masses based on ALMA continuum observations of gaps.
\end{abstract}


\section{Introduction}
There is no unambiguous detection of a circumplanetary disk (CPD) formed around young gas-giants. However several traces have been found, for instance, H-alpha emission from the planetary candidate LkCa15b \citep{Sallum15} was detected. Traces of warm excess emission around planet candidates at near-infrared wavelengths have been observed in a number of cases, for example, LkCa15b \citep{KI12}, two planetary candidates in the disk of HD100546b \citep{Quanz15,Britain14}, the companion(s) of HD169142 \citep{Reggiani14,Osorio14}. Detecting and characterizing CPDs could get us closer to understand satellite formation, but also to help distinguishing between the different planet formation paradigms, such as core accretion versus disk instability, or cold-start versus hot-start formation scenarios \citep[e.g.][]{Marley07}. First of all, according to \citet{SzM16} the CPD changes the entropy of the accreted gas that can alter the entropy of the planet. Secondly, the bulk CPD temperature could be significantly higher around core-accretion planets than disk instability formed ones \citep{Szulagyi16b}.

Naturally, hunting for CPDs has begun with the Atacama Large Millimeter Array (ALMA) with several ongoing projects, with so far only upper limits for the mass of the sub-disk (e.g. Pineda et al. in prep.). In the case of a brown dwarf in the system of GSC 6214-210, ALMA could set an upper limit on the circum-substellar disk mass \citep{Bowler15}. Mock observations by \citet{Perez15} predict that by pushing ALMA to its sensitivity limits the kinematic imprint of the rotating CPD can be spotted within the surrounding circumstellar disk. 

Both the core-accretion and the disk instability planet formation mechanisms lead to CPDs around the young planet \citep{Szulagyi16b}. \citet{Mayer16} used mock observations to show that collapsing protoplanets via gravitational instability can also be observed with ALMA, however their circumplanetary disks are unlikely to be resolved. The CPD fills radially only 30-50\% of the Hill-sphere \citep[e.g.][]{Tanigawa12,Szulagyi14}, therefore even massive planets at large orbital distances have quite small sub-disks for ALMA's spatial resolution. For instance, a 10 Jupiter-mass planet at 50 AU around a Solar-mass star would have a Hill-sphere radius of $\sim$7.5 AU, that means a maximal CPD size of only $\sim$3.7 AU. But even if resolving the CPD is out of scope, detecting it might still be possible with ALMA as it is investigated in this paper.

Forming giant planets significantly interact with their circumstellar disk, creating gaps and spirals. Several works in the recent years focused on creating mock observations of this planet-disk interaction, many of which had predictions for ALMA. Models which included dust coagulation \citep{Pinilla2015a,Pinilla2015b} concentrated on the dust trapping by the planetary gap edges and the resulting ring structures which are seen with ALMA in numerous circumstellar disks. With a two-fluid (gas+dust) approach for hydrodynamic simulations, the planet opened gaps were studied in \citet{Dong15}. The differences of dust and gas gaps in ALMA real and mock observations were highlighted in \citet{Dong17} on a case study for the transitional disk J160421.7-213028. The study of \citet{IsellaTurner16} showed that the vertical temperature profile of the circumstellar disk can significantly alter the outcome of how the planet-disk interaction looks like in synthetic observations.

Even though previous hydrodynamical simulations often included some dust treatment, they were often restricted to 2D and/or without including any realistic thermal effects. With the lack of a proper heating \& cooling, the calculated fluxes and structures triggered by forming planets cannot be fully investigated. Therefore, here we present three dimensional, radiative hydrodynamic simulations of forming gas-giants embedded in circumstellar disks. The radiative module in the hydrodynamic code includes radiative cooling, viscous- and shock-heating and adiabatic heating/cooling. These effects significantly change the temperature in the planet vicinity. Therefore, with this radiative module, the imprint of accreting planet can be more realistically studied in ALMA mock observations than in previous works.

The main heating mechanism in the CPD is the adiabatic compression due to the accretion process \citep{Szulagyi16a}. The viscous heating is small, because the ionization rate is very low in the bulk of the CPD \citep{Fujii11,Fujii14,Fujii17}. The thermal ionization is only efficient in the very inner part of the disk, that touches the planet. The shock on the surface of the CPD that is created from the vertical influx of gas from the circumstellar disk, can also ionize locally, but only in razor sharp surface \citep{SzM16}. Besides, the stellar photons cannot reach and ionize the CPD, as long as the inner circumstellar disk is present, because the scaleheight of the CPD is significantly smaller, therefore shadowed by the circumstellar disk. In summary, the CPD has to have small viscosity, and it is not acting like an $\alpha$-disk \citep{Sunayev}, hence cannot be modeled as such.

This work is the first in a series where the observability of CPDs is explored at various wavelengths. Follow-up works will study the question in near-infrared and in scattered light.

\section[]{Methods}
\label{sec:numerical}

In this paper, synthetic images of forming gas giant planets is presented that can be found within protoplanetary disks, concentrating on the observability of the CPDs with ALMA. Radiative hydrodynamic simulations \citep{Szulagyi17} were carried out, which were post-processed with RADMC3D, a wavelength-dependent radiative transfer tool, as well as the ALMA simulator in order to obtain the final product. In Sect. \ref{sec:hydro} the basic parameters of the hydrodynamic simulations are summarized, then in Sect. \ref{sec:radmc3d} the radiative transfer assumptions are described, finally, in Sect. \ref{sec:almasim} the ALMA Simulator input values are discussed.

\subsection{Hydrodynamic Simulations}
\label{sec:hydro}

The base models are radiative, three-dimensional hydrodynamic simulations of forming planets in a circumstellar disk. For these the JUPITER code was used, that was developed by F. Masset and J. Szul\'agyi \citep{Szulagyi14,Szulagyi16a,Borro06}. Apart from solving the basic hydrodynamical equations, it also solves the total energy equation and contains flux-limited diffusion approximation with the two-temperature approach \citep[e.g.][]{Commercon11,Bitsch14,Kley89}. This radiative module takes care of the inclusion of thermal effects: the gas can heat due to adiabatic compression and viscous heating, while it cools through radiation and adiabatic expansion. The accretion onto the planet increases the adiabatic compression in the circumplanetary disk near the planet, which is the main heating mechanism in the gas-giant's vicinity (see also in \citealt{Szulagyi16a}, and in \citealt{Montesinos15}). More specifically, the radiative flux
\begin{equation}
\mathbf{F_{rad}}=-\frac{c\lambda}{\rho \kappa_R} \nabla \cdot \epsilon_{rad} 
\end{equation}
is given by the flux-limited diffusion approximation \citep{Levermore81}, where $\epsilon_{rad}$ is the radiative energy, $\rho$ represents the density, $\kappa_R$ is the mean Rosseland-opacity, and $c$ stands for the speed of light. $\lambda$ denotes the flux limiter that reduces the flux in the following way: it approaches to $F=4\sigma T^4/c$ ($\sigma$ is the Stefan-Boltzmann constant and $T$ represents the temperature) in the optically thin parts, while it approaches $\lambda = 1/3$ in the optically thick parts. Therefore $\lambda$ accounts for the smooth transition between the optically thick and thin regimes. The flux-limiter was defined according to \citet{Kley89} and \citet{Kley09}.

For the temperature calculation, the opacity tables of \citet{BL94} were used that contain both gas and dust opacities. Therefore, even though there is no explicit dust treatment in the simulations, the dust contribution to the temperature is taken into account with a constant dust-to-gas ratio of 0.01. Due to the combined dust and gas opacity table, the gas will provide the opacity above the dust sublimation point ($>$1500 K).

The simulations used in this work were partially published in \citet{Szulagyi17} and \citet{Szulagyi16a}, therefore, here only their most important characteristics are summarized. The JUPITER code also includes mesh refinement, so that the resolution is enhanced in the planet's vicinity, reaching a peak resolution of 80\% of Jupiter's diameter, that is approx. 110000 km. Nevertheless, the simulations were global, i.e. they contain a radially extended ring of the circumstellar disk between 2.1 and 12.5 AU with the planets placed at 5.2 AU. Even though real protoplanetary disks can extend beyond 100 AU, the outer disk beyond 12 AU has no effect on the CPD region, but simulating such an extended disk would increase significantly the computation time. The protoplanetary disk had a mass of $\sim 11 \mathrm{M_{Jup}}$ with a surface density slope of 0.5 initially, which evolved under the heating-cooling effects and the inclusion of the high mass planets. Each simulation contained one planet, whose masses were either Saturn, Jupiter, $3\mathrm{M_{Jup}}$, $5\mathrm{M_{Jup}}$, or $10\mathrm{M_{Jup}}$. The CPD temperature is also affected by the planet temperature \citep{Szulagyi16a} so an assumption has to be made regarding this value. The effective temperatures of forming planets are poorly constrained in the non-detached phase even with evolution models. Planet interior models earliest prediction is at 1Myr of age, in most cases assuming that the planets form in vacuum that result in a temperature underestimation, that is 1000 K for Jupiter \citep[e.g.][]{Guillot95}. In the most recent planet population simulations with interior models that contain accretional luminosity and a background disk \citep{Mordasini17}, the effective temperatures of planets with a few Jupiter-masses range between 2000-8000 K at 1 Myr. Therefore in most of our simulations the planet had a fixed temperature within 3 Jupiter-radii to a middle-ground-value of 4000 K. To test the effect of the planet temperature on the resulting ALMA fluxes, a simulation of a Jupiter-mass planet with only 1000 K surface temperature was also carried out.

The equation-of-state in all computations was ideal gas: $P=(\gamma-1)e$, where $\gamma=1.43$ adiabatic index connects the pressure $P$ with the internal energy $e$. A constant kinematic viscosity of $10^{-5} \mathrm{a_{p}}^2\Omega_p$ was applied, where $ \mathrm{a_{p}}$ denotes the semi-major axis and $\Omega_p$ represents the orbital frequency of the planet. This corresponds to an $\alpha$-viscosity of $\sim 0.004$ at the planet's location.  The mean molecular weight was kept to a value of 2.3 representing the solar abundances.

The different simulations were run the same way, introducing the refined grid levels at the same time, after steady state has been reached on each level. The simulations results were obtained after the 240th orbit of the planet. 


Our goal in this work is to understand in what conditions the CPDs can be detected with ALMA and why previous attempts were not successful. Furthermore, if the detection is successful, what can that tell about the CPD and the planet itself. This study is aiming at helping the community to detect the imprints of forming planets and characterize them along with their disks.

\subsection{RADMC3D post-processing}
\label{sec:radmc3d}

The hydrodynamic simulations were radiative, but wavelength independent. They used a Rosseland mean opacity, therefore to obtain the continuum intensity images at a given wavelength (in a given ALMA Band), the RADMC3D \citep{Dullemond12}\footnote{\url{http://www.ita.uni-heidelberg.de/~dullemond/software/radmc-3d/}} radiative transfer tool was used.

First, the grid system of the JUPITER code was transformed into RADMC3D conventions with a self written IDL pipeline. Due to some inner boundary condition effect, the inner 26 cells off from the circumstellar disk (up until 3 AU from the star) were cut off. For the radiative transfer input parameters, a $1.0 \rm{M_{\odot}}$ star with $\rm{T_{eff}}=5800$ K and $1.0 \rm{R_{\odot}}$ was assumed in order to be consistent with the hydrodynamic model assumptions. Due to the 3D spherical coordinates of the models, the star was treated as a sphere. For the disk temperature, we used the (gas) temperature calculated by the hydrodynamic code and assumed that the dust temperature was equal to this (perfect thermal equilibrium). The reason for this was that in this work we are particularly interested in the circumplanetary disk and RADMC3D's Monte-Carlo approach calculating the dust temperatures are too simplistic in this case. During a test it was found that the RADMC3D's Monte-Carlo approach leads to a significantly lower dust temperature in the planet's vicinity, hence the image barely shows the presence of the planet and its circumplanetary disk. The newest version of RADMC3D, v0.41, includes addition of a heatsource with definining the additional energy source at each cell, however to figure out the combined contribution of adabatic heating due to accretion, viscous heating and the cooling mechanisms, a hydrodynamic simulation is probably needed for a precise determination of energy per cell. Although the hydrodynamic code temperature calculation is also far from being perfect (e.g. neglects ionization of the gas), it still gives more reasonable values for the temperature in the planet's vicinity than a simple Monte Carlo approach. As it will be discussed later (in Sect. \ref{sec:res}), the CPD is optically thick in most ALMA bands studied in this paper, therefore the $\rm{T_{dust}=T_{gas}}$ assumption is acceptable. Regarding the dust density distribution, with the lack of a dust treatment in the hydro models, we assumed the distribution to be the same as for the gas, but only 1\% in mass fraction (again to be consistent with the initial parameters of the hydrodynamic simulations). It was checked whether the strong coupling between the dust and gas has been a valid assumption by calculating the Stokes number (using a grain size of 1 mm, the density of dust particles of 3 $\rm{g/cm^3}$, and the computed gas surface density from the simulations). If the Stokes number is less than unity, then strong coupling can be assumed. Our test revealed that the Stokes number is less than 0.17 everywhere in our simulation box, hence the coupling between dust and gas is indeed strong.

Only one dust species was used for the dust opacity table, a mixture of silicates \citep{Draine03}, carbon \citep{Zubko96} with fractional abundances of 70\% and 30\%. The opacity of the mixture was determined by the Bruggeman mixing formula. The absorption and scattering opacities, as well as the scattering matrix elements were calculated with Mie theory considering the BHMIE code of \citet{BH84}. The dust grain size distribution ranged between 0.1 micron and 1 cm according to a power-law with index of -3.5, similarly to \citet{Pohl17}. The absorption opacity values at 350, 440, 740, 870, 1300, and 2100 microns are 17.1, 15.4, 11.5, 10.2, 7.8, 5.6 $\rm{cm^2 g^{-1}}$, respectively.

The RADMC3D model resolution were set to 1000 $\times$ 1000 pixels in each case to avoid resolution problems. The distance was assumed to be 100 parsec. 

\subsection{ALMA Simulator}
\label{sec:almasim}

The RADMC3D intensity maps were processed with the Common Astronomy Software Applications' (CASA) \texttt{simobserve} \& \texttt{simanalyze} tools to create mock dust continuum observations. The full, 50-antenna ALMA was considered (i.e. even those antennas that will be placed in the near future) with the most extended configuration (\#28) to achieve the highest resolution, that is 0.005 arcseconds for the shortest wavelength bands. This resolution with 100 pc distance means a 0.5 AU resolution, that is still much larger than the entire Hill-sphere when the planet is at 5.2 AU. As it was mentioned before, the CPD is a subset of the Hill-sphere (approx. 50\% $\rm{R_{Hill}}$, e.g. \citealt{Szulagyi14}, \citealt{SB13}, \citealt{AB09}), therefore resolving the Hill-sphere for any of our planetary masses is not possible. However, when the planet has significantly larger orbital separation from its star, the angular resolution can be decreased accordingly. See a test about this in Sect. \ref{sec:orbital_sep}.

The integration time was chosen to be 10 seconds for each pointing with a total integration time of one hour, except for the planet at 52 AU, where 3 hours of total time was required for a detection (Sect. \ref{sec:orbital_sep}). Note, that this integration time is not representative, because it is not equal to the sensitivity calculator's value to reach the same sensitivity. According to the manual, this is a known issue, and the value given in the sensitivity calculator should be taken for a proposal and the discrepancy between the simulator and the sensitivity calculator tools is increasing toward the shorter wavelength bands. According to our tests, e.g. in band 9, one needs approx. five hours total integration time according to the sensitivity calculator to reach the same sensitivity as the simulator gives with one hour integration time. The bandwidths of continuum observations were 7.5 GHz for the bands 8, 7, 6, 4 and 15 GHz in the case of Band 9 and 10. This is because since Cycle 5 the sensitivity of band 9 and 10 is $\sqrt{2}$ higher than in the previous years due to the Walsch switchers. The theoretical RMS values reached with the simulator, as well as the asked frequencies for the continuum can be found in Table \ref{tab:mock_jup3}.

Noise was added with the \texttt{corrupt} command, including the phase noise that mainly affects the shortest wavelength bands (band 9 \& 10). The ground temperature was assumed to be 269 K, altitude of 5000 meters, humidity of 20\% and atmospheric condition of 0.475 mm water. These are optimistic weather conditions, however, short wavelengths observations do require such conditions and will be carried out only in exceptional weather. In addition, the receiver temperatures (``trx" values) were 230 K, 110 K, 196 K, 75 K, 55 K, 51 K for band 10, 9, 8, 7, 6, 4 respectively, according to the user manual of ALMA. The command to add the noise was the following:         

\begin{verbatim}
sm.setnoise(mode='tsys-atm',pground='560mbar',
altitude='5000m',waterheight='200m',relhum=20,
pwv="0.475mm",spillefficiency=0.85,
correfficiency=0.845,antefficiency=0.8,trx=trxarr[xx],
rxtype=1,tground=269.0,tatmos=250.0,tcmb=2.725) 

sm.settrop(mode='screen', pwv=0.475, deltapwv=0.1)
\end{verbatim}
    
\section{Results}

\subsection{Circumplanetary Disk Observational Predictions}
\label{sec:res}

\begin{figure}
\includegraphics[width=\columnwidth]{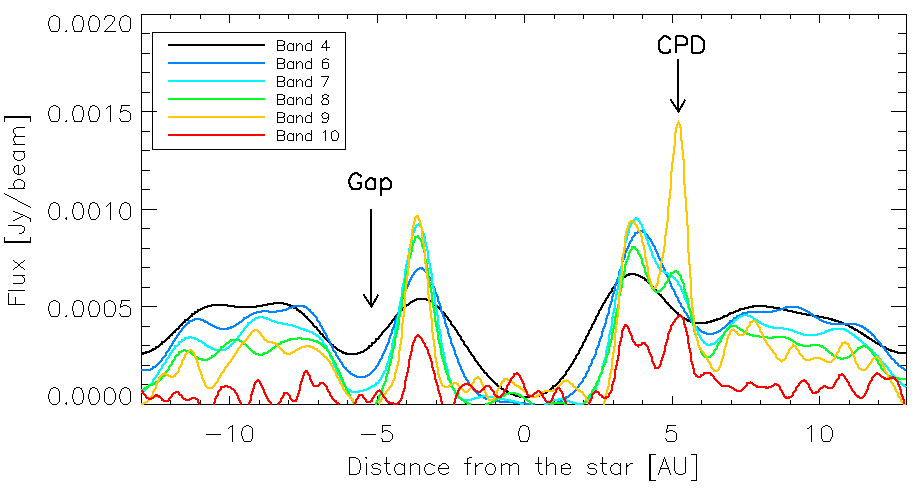}
\caption{Slice through the mock observations of the 3 Jupiter-mass planet at various ALMA Bands, showing the CPD contribution and the planetary gap profiles. The CPD pops out from the background circumstellar disk only on Band 9 and 10.}
\label{fig:CPD_over_gap}
\end{figure}

We created continuum mock observations of the different planetary mass simulations. Table \ref{tab:mock_jup3} shows the 3 Jupiter-mass planet mock observations at Band 10, 9, 8, 7, 6, and 4. According the images in this table, the detection of the hot spot CPD is more prominent in Band 9 or 10 than at longer wavelengths. To compare the the different bands with each other, a slice was made through the planet location and plotted this in Fig. \ref{fig:CPD_over_gap}. It can be seen that the CPD pops out from the rest of the circumstellar disk in Band 9 and 10.

\begin{table*}
  \caption{Mock observations for the different ALMA Bands of the 3 Jupiter-mass hydro-simulation}
 \label{tab:mock_jup3}
  \begin{tabular}{ccc}
  \hline
Band 10 & Band 9  & Band 8\\
(350 microns, 856.5 GHz) &(440 microns, 681 GHz) & (740 microns, 405 GHz)\\
 $\rm{RMS_{theoretical}}=3.2\times10^{-4}$ Jy &  $\rm{RMS_{theoretical}}=1.3\times10^{-4}$ Jy &  $\rm{RMS_{theoretical}}=3.8\times10^{-5}$ Jy  \\
$\rm{RMS_{measured}}=5.5\times10^{-5}$ Jy & $\rm{RMS_{measured}}=3.1\times10^{-5}$ Jy & $\rm{RMS_{measured}}=2.4\times10^{-5}$Jy  \\ 
BMAJ = 0.006", BMIN = 0.006" & BMAJ = 0.008", BMIN = 0.007" & BMAJ = 0.013", BMIN = 0.012"\\
 \hline
   \includegraphics[width=0.7\columnwidth]{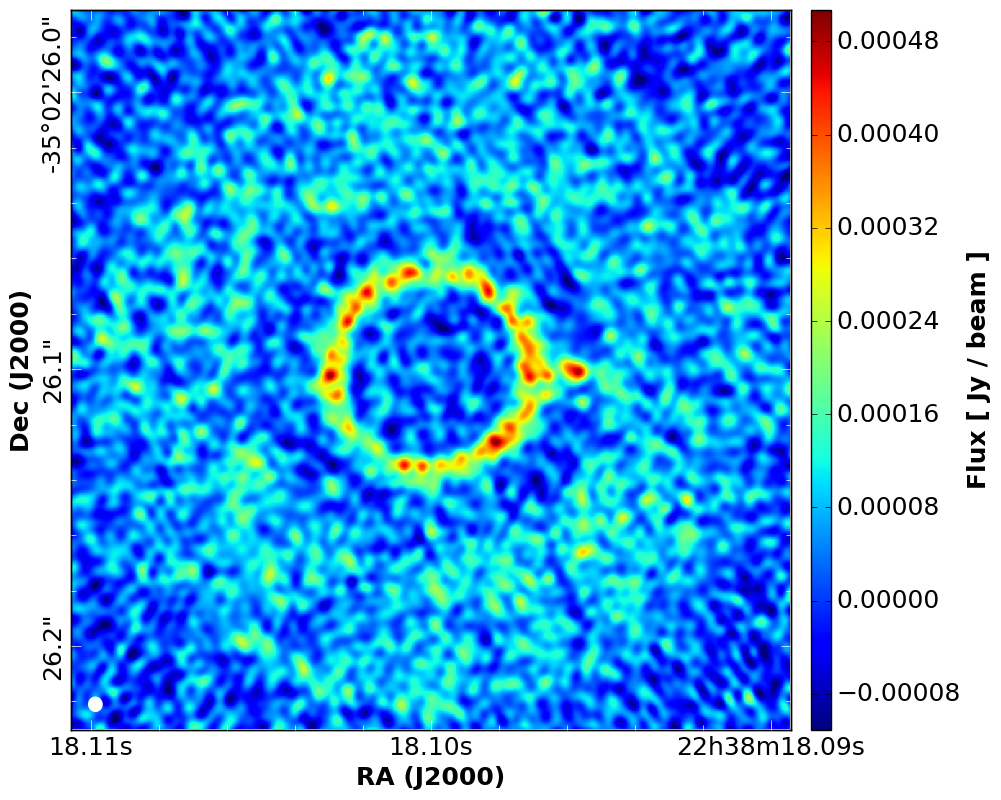}	&	\includegraphics[width=0.7\columnwidth]{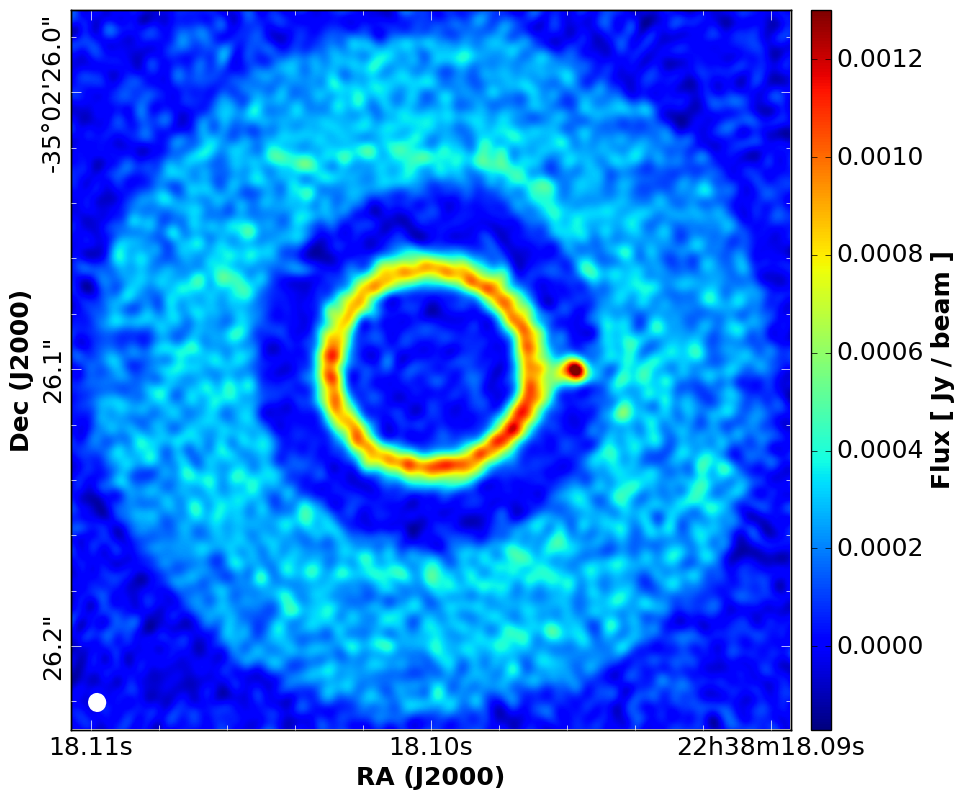}	&	\includegraphics[width=0.7\columnwidth]{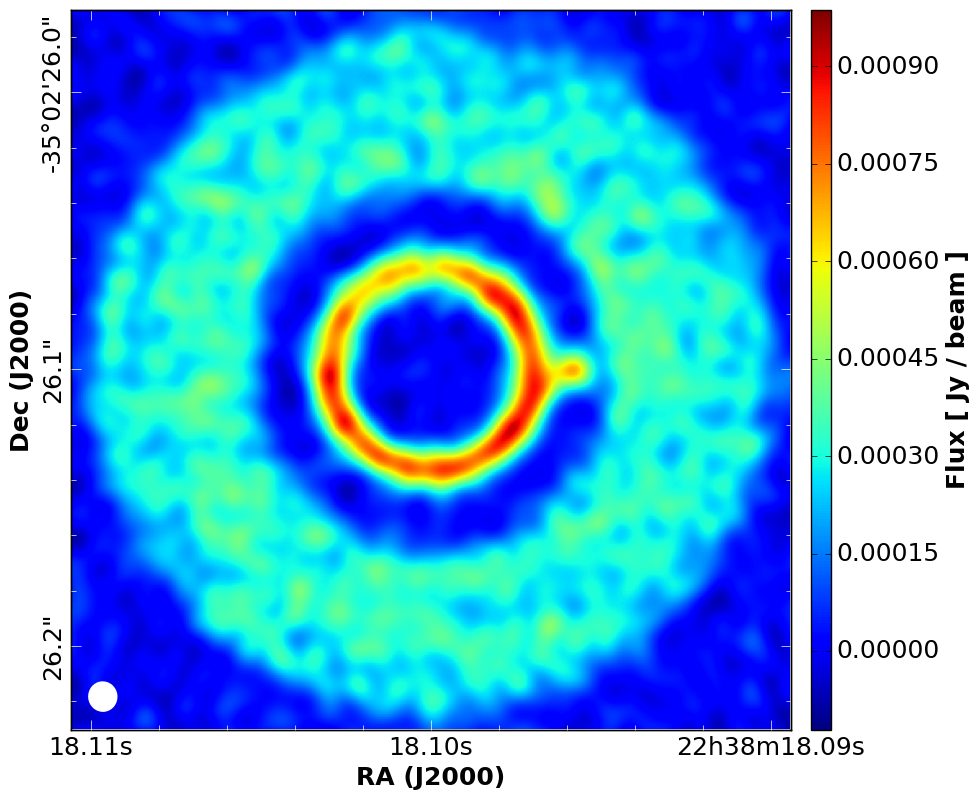}	\\
  \hline
 Band 7 & Band 6  & Band 4 \\
(870 microns, 344.5 GHz) & (1300 microns, 230 GHz) & (2100 microns, 142.8 GHz) \\
$\rm{RMS_{theoretical}}=1.4\times10^{-5}$ Jy & $\rm{RMS_{theoretical}}=8.8\times10^{-5}$ Jy  & $\rm{RMS_{theoretical}}=7.6\times10^{-5}$ Jy \\
$\rm{RMS_{measured}}=1.4\times10^{-5}$ Jy & $\rm{RMS_{measured}}=1.8\times10^{-5}$ Jy & $\rm{RMS_{measured}}=1.3\times10^{-5}$ Jy \\
BMAJ = 0.015", BMIN = 0.015" & BMAJ = 0.023", BMIN = 0.022" & BMAJ = 0.038", BMIN = 0.036"\\
  \hline
\includegraphics[width=0.7\columnwidth]{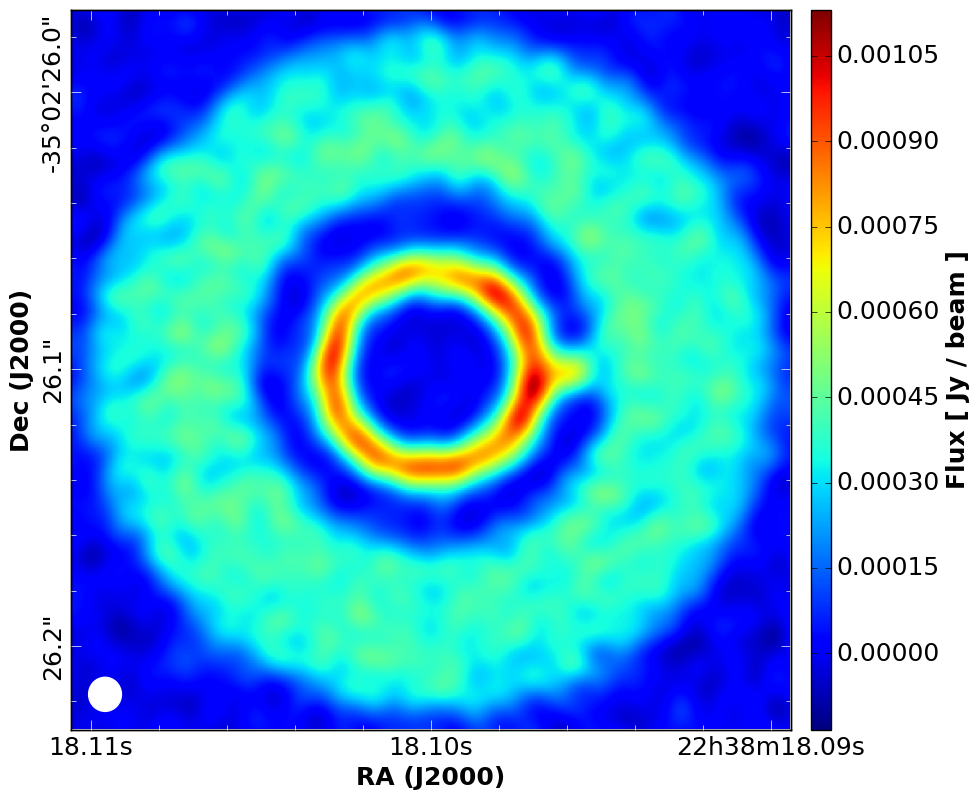} &\includegraphics[width=0.7\columnwidth]{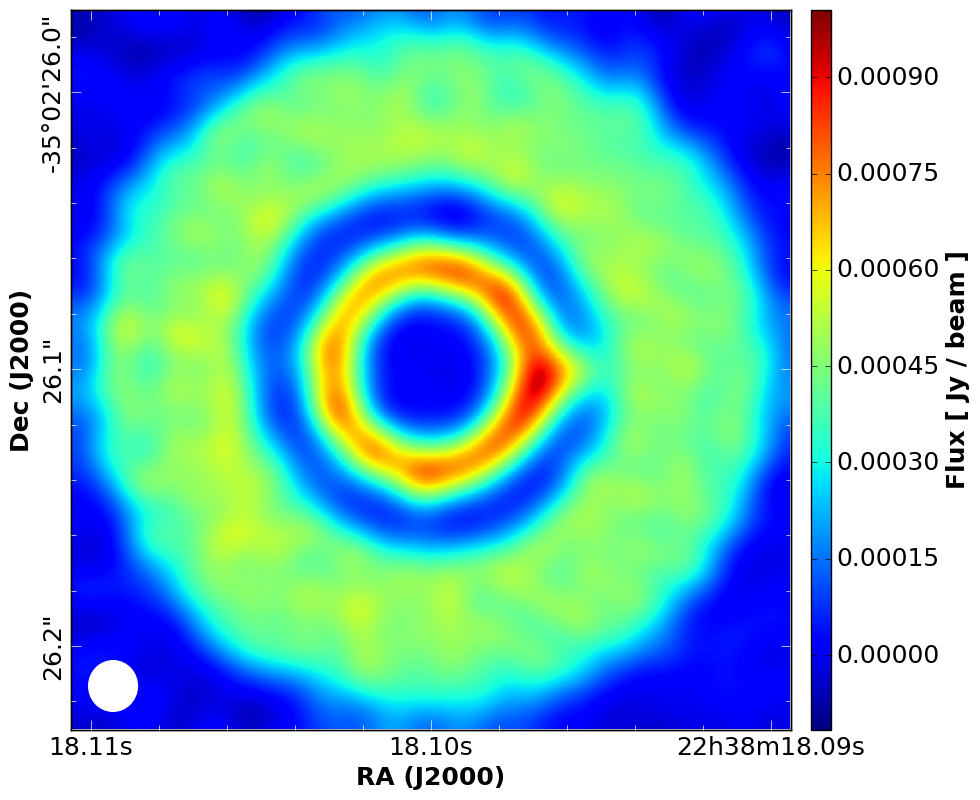} & \includegraphics[width=0.7\columnwidth]{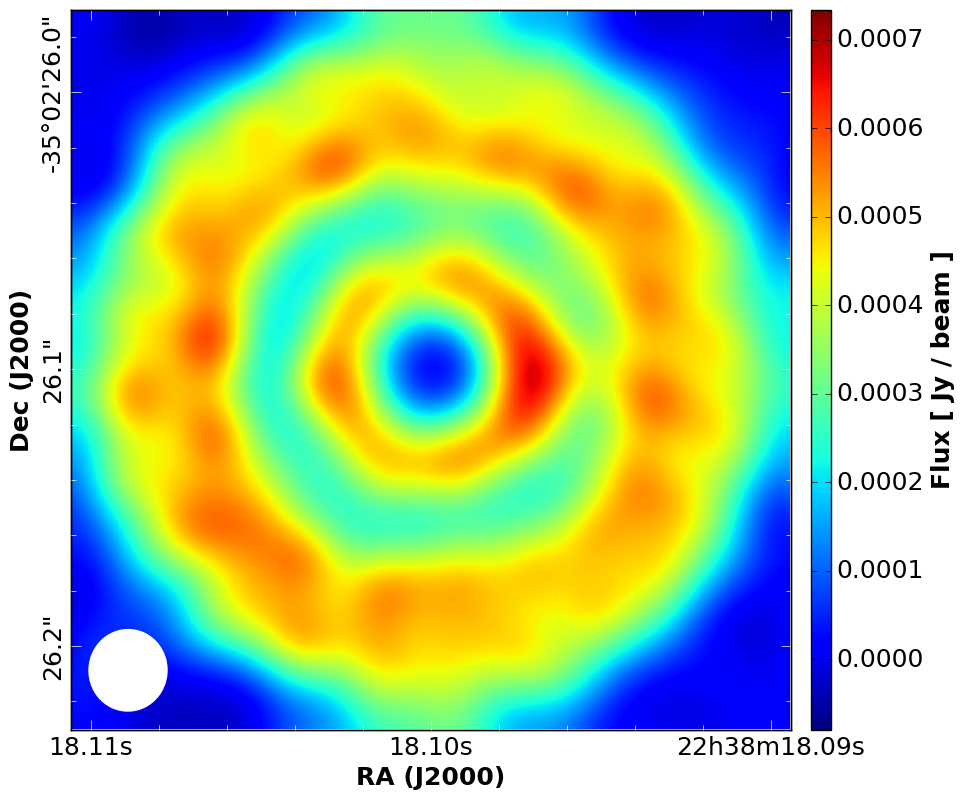}\\
\end{tabular}
\end{table*}

\begin{figure*}
\includegraphics[width=17cm]{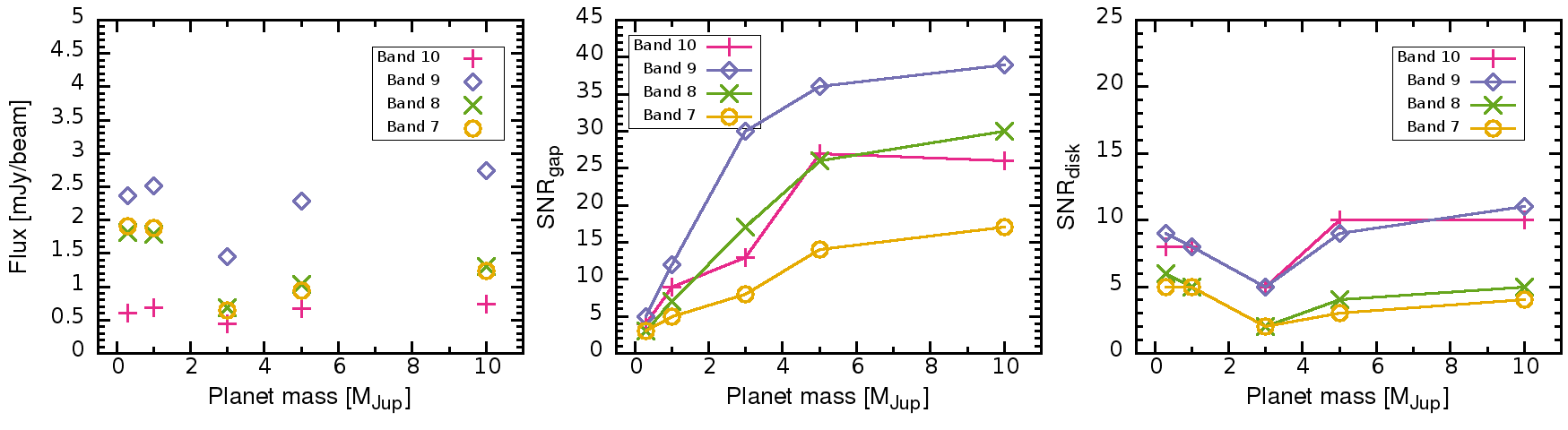}
\caption{Left: ALMA continuum peak fluxes at the planet location for Bands 10, 9, 8, and 7 (350, 440, 740, 870 microns, respectively). Middle: SNR relative to the gap which increases with planetary mass as the gaps get wider and deeper. Right: SNR relative to the circumstellar disk showing that a Saturn mass planet is almost as easy to detect as a 10 Jupiter-mass planet.}
\label{fig:flux_cap}
\end{figure*}

The peak fluxes at the planet's location were measured in each simulation in Band 10, 9, 8, 7, 6, and 4 for the different planetary masses (see Table \ref{tab:fluxes}, left panel in Fig. \ref{fig:flux_cap}). Because the Saturn mass planet did not open a deep gap, its CPD-flux was contaminated by the background circumstellar disk flux, therefore in this case, the measured CSD flux was subtracted by using the location at the same distance from the star but on the opposite side where the planet lies. From the numbers it is clear that in Band 9 (440 microns) one obtains higher fluxes on the CPD for each planetary masses than in the longer wavelength bands (left panel on Fig. \ref{fig:flux_cap}). This is due to the fact that the circumplanetary disk is hot in the optically thick limit during the gaseous circumstellar disk phase due to the accreting proto-planet and the slow cooling time. Using the Sensitivity Calculator tool, an integration time of 5 hours was calculated to achieve the reported sensitivity in band 9 with the new (from Cycle 5) 15 GHz bandwidth.

\begin{table*}
  \caption{Circumplanetary Disk Peak Fluxes and Signal-to-Noise Ratios}
 \label{tab:fluxes}
  \begin{adjustbox}{max width=\textwidth}
  \begin{tabular}{l|ccc|ccc|ccc|ccc|ccc|ccc}
  \hline
& Band 10 &&& Band 9 &&& Band 8  &&& Band 7 &&& Band 6  &&& Band 4 && \\
  \hline
 $\rm{M_{P}}$ & $F_{\nu}$ & $\rm{SNR_{gap}}$ & $\rm{SNR_{disk}}$ &  $F_{\nu}$  & $\rm{SNR_{gap}}$ & $\rm{SNR_{disk}}$&  $F_{\nu}$  & $\rm{SNR_{gap}}$ & $\rm{SNR_{disk}}$& $F_{\nu}$  & $\rm{SNR_{gap}}$ & $\rm{SNR_{disk}}$& $F_{\nu}$  & $\rm{SNR_{gap}}$ & $\rm{SNR_{disk}}$& $F_{\nu}$  & $\rm{SNR_{gap}}$ & $\rm{SNR_{disk}}$\\
 $[\rm{M_{Jup}}]$ &   [mJy/beam] &  &  & [mJy/beam]&  &  &  [mJy/beam] &  &  &   [mJy/beam] &  &  &   [mJy/beam] &  & &   [mJy/beam] &  &   \\
 \hline
0.3 & 0.60 &            4 &            8 &  2.36 &            5 &            9 &  1.82 &            3 &            6 &  1.91 &            3 &            5 &  1.48 &            2 &            3 &  0.84 &            1 &            2 \\
1.0 &  0.69 &            9 &            8 &  2.52 &           12 &            8 &  1.79 &            7 &            5 &  1.88 &            5 &            5 &  1.50 &            3 &            3 &  1.02 &            2 &            2 \\
3.0 & 0.45 &           13 &            5 &  1.45 &           30 &            5 &  0.68 &           17 &            2 &  0.65 &            8 &            2 &  0.60 &            3 &            1 &  0.50 &            2 &            1 \\
5.0 & 0.67 &           27 &           10 &  2.28 &           36 &            9 &  1.04 &           26 &            4 &  0.94 &           14 &            3 &  0.68 &            6 &            2 &  0.47 &            2 &            1 \\
10.0 & 0.74 &           26 &           10 &  2.74 &           39 &           11 &  1.31 &           30 &            5 &  1.23 &           17 &            4 &  0.86 &            9 &            2 &  0.51 &            3 &            1 \\
\hline
\multicolumn{19}{l}{{\it Note:} In the case of the Saturn mass (0.3 $\rm{M_{Jup}}$) planet, the CPD sits on a background CSD, due to the imperfect gap-opening, therefore the CSD contribution to the CPD flux was subtracted. In}\\
\multicolumn{19}{l}{the other cases the CPD sits in the gap, therefore such correction was not needed.}
 \end{tabular}}
 \end{adjustbox}
 \end{table*}

One might expect to have a flux dependence with the planetary masses: the higher the planetary mass, the higher the temperature, so a priori, the higher is the luminosity. However, the left panel in Fig. \ref{fig:flux_cap} shows that one can expect similar fluxes from a Saturn or a 1 Jupiter-mass planet as from a 10 Jupiter-mass gas giant. There are multiple effects playing a role here: gap-opening and the density and temperature (which affects the optical depth at a given wavelength) in the planet vicinity. First, Saturn opens only a partial gap and forms a tiny CPD that is well surrounded by the circumstellar disk. Because it does not clear up its orbit as much as the higher mass planets, there is a significant amount of gas and dust in planet's surroundings, which can therefore cool less efficiently. In other words, less massive forming planets, which cannot open wide gas gaps are heating the surrounding circumstellar disk, therefore their heating effects are affecting a larger area. On the other hand, the wider is the gap, the less the circumplanetary gas and dust is blanketed, and the radiation can quickly escape. This can be easily understood from the optical depth maps in Table \ref{tab:tau_maps}, where the optically thick ($\tau > 1.0$) regions of the circumstellar disks are displayed. These maps were created by calculating vertically integrated optical depth from the density and Rosseland mean opacity from the hydrodynamic simulations. The flux versus planetary-mass relation (left panel on Fig. \ref{fig:flux_cap}), however, means that it would be difficult to estimate the planetary masses from the subm-mm observations, since the relation is not linear and highly dependent on the planet temperature. To detect the CPD with ALMA, it seems that Saturn or a Jupiter-mass planet is a better target based on the fluxes than most of the larger mass gas-giants. 

\begin{table*}
  \caption{Optically thick regions of the circumstellar disks}
 \label{tab:tau_maps}
  \begin{tabular}{ccc}
  \hline
Saturn & 1$\mathrm{M_{Jup}}$   & 3$\mathrm{M_{Jup}}$\\
 \hline
   \includegraphics[width=0.7\columnwidth]{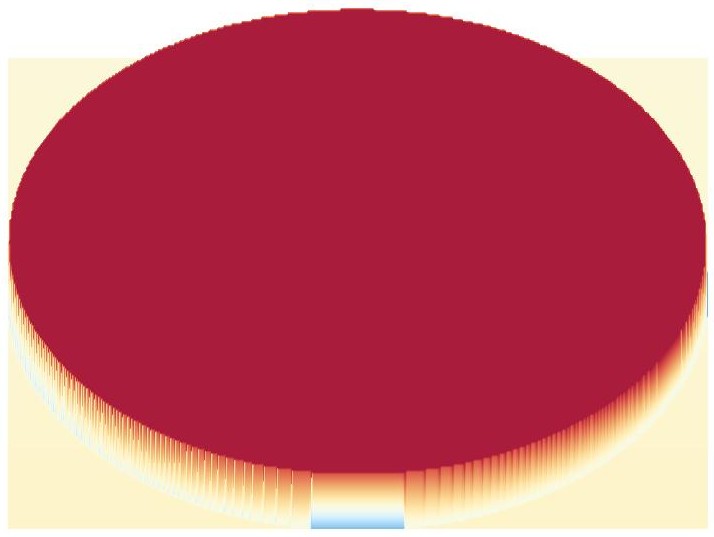}	&	\includegraphics[width=0.7\columnwidth]{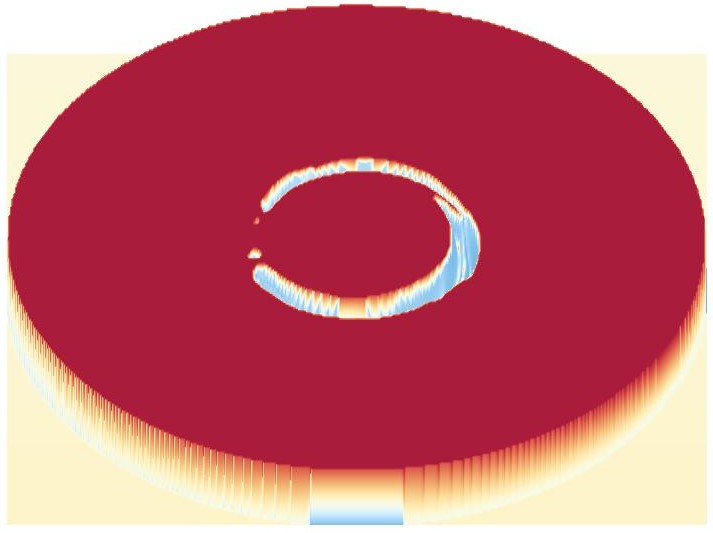}	&	\includegraphics[width=0.7\columnwidth]{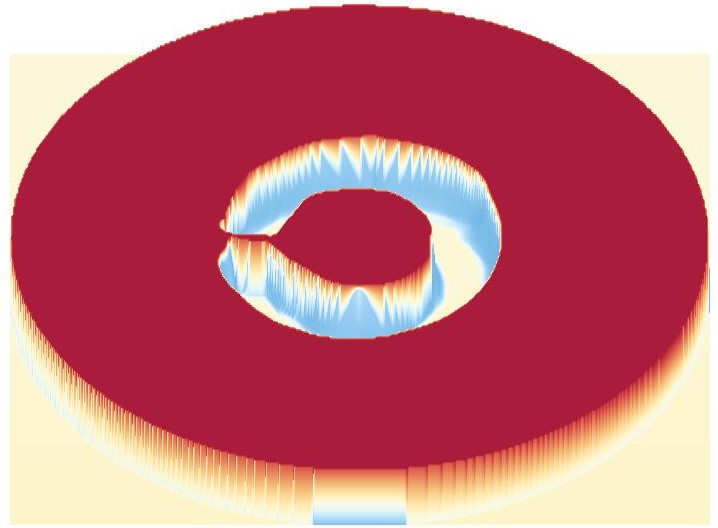}	\\
  \hline
 5$\mathrm{M_{Jup}}$ & 10$\mathrm{M_{Jup}}$ & \\
  \hline
\includegraphics[width=0.7\columnwidth]{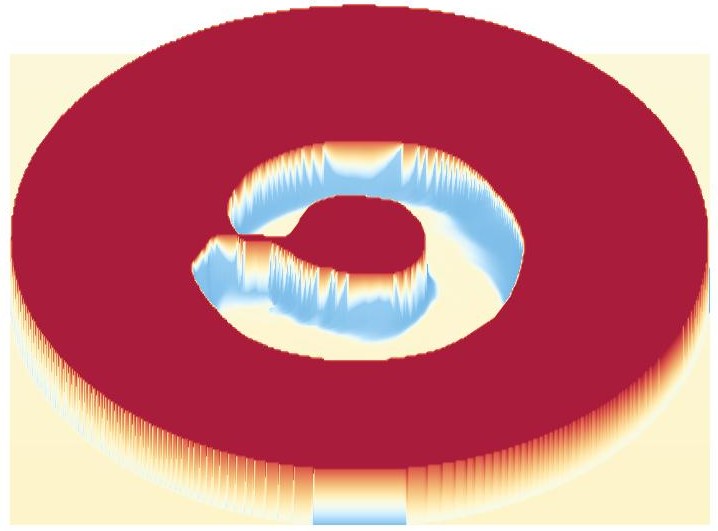} &\includegraphics[width=0.7\columnwidth]{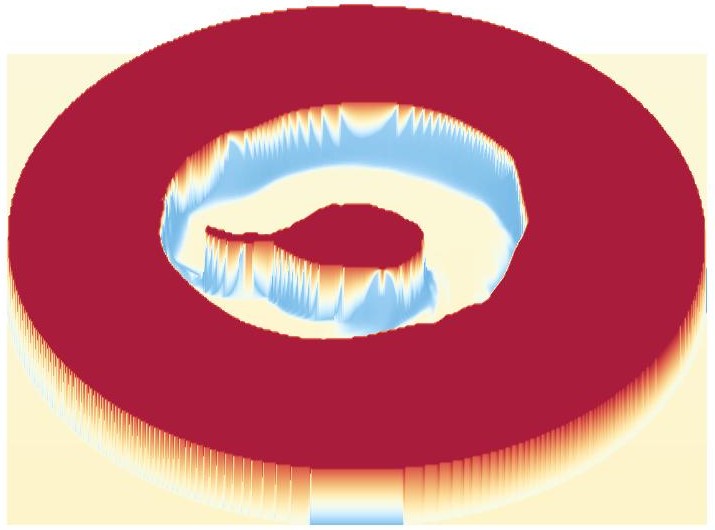} & \\
\end{tabular}
\end{table*}

The root-mean-square (RMS) of the synthetic images are shown in Table \ref{tab:mock_jup3}. The theoretical noise limits are from the web-based ALMA simulator \footnote{http://almaost.jb.man.ac.uk/} that is calculated on the naturally weighted images. The measured RMS is calculated on these CASA produced synthetic maps (that are Briggs weighted) with IDL using the lower left corners of each image outside the circumstellar disk. The reason that the theoretical and measured RMS is different is a known issue in the ALMA manual, partially it comes from the fact that different noise components are taken into acocunt in one and in the other. To quantify the CPD detection capability with a signal-to-noise ratio (SNR), one could use the noise limits listed in Table \ref{tab:mock_jup3}. However, this way the SNR values would be naturally lower for longer wavelengths just because of the larger beam size, and the same integration time used for all bands (Sect. \ref{sec:almasim}). The synthetic images for Band 9 at the various planetary masses (Table \ref{tab:mock_sequence}), however, show that the CPD detectability depend mostly on the CPD contrast relative to the gap and the circumstellar disk. Therefore, our $\rm{SNR_{gap}}$ was calculated by dividing the CPD peak flux value with an azimuthally averaged gap flux value. The middle panel in Fig. \ref{fig:flux_cap} shows that there is a trend of increasing $\rm{SNR_{gap}}$ with growing planetary masses, as these gas-giants open yet deeper and wider gaps. Based on the contrast with the gap, the 5 and 10 $\rm{M_{Jup}}$ planets are the best targets for CPD hunting. However, the images in Table \ref{tab:mock_sequence} show that the contrast with the circumstellar disk is even more important for CPD detectability. Hence the second SNR value reported in Table \ref{tab:fluxes} defines this by dividing the CPD peak flux value by an azimuthally and radially averaged disk profile of the circumstellar disk (eliminating radially the gaps and CPDs from the average). The right panel in Fig. \ref{fig:flux_cap} displays $\rm{SNR_{disk}}$, revealing that Saturn-mass planets are almost as good targets to detect the hot, circumplanetary dust than the 10 $\rm{M_{Jup}}$ gas giants. According to the synthetic images in Table \ref{tab:mock_sequence}, the imprint of the Saturn-mass giant is even easier to see than of the 10 Jupiter-mass planet's. One has to be careful though with the interpretation whether the Saturn case qualifies as a CPD detection, because, as it was mentioned before, Saturn-mass planets open only partial gaps, which are still significantly filled with gas and their CPD is tiny (in radial extent). Therefore, here the observed flux is not that much from the hot dust of the CPD but it is combined with the hot circumstellar disk gas+dust around the CPD. Further testing is needed with even lower mass planets than Saturn to see whether this trend is the same, when there is no CPD, but a planetary envelope that is directly surrounded by the circumstellar disk and there is not even partial gap-opening. In the likely case that the result will be similar to our Saturn case, it means that an extended sub-mm emission around a point source in a circumstellar disk does not mean a presence of a CPD, only a presence of a forming low-mass planet that heats its surrounding circumstellar material. Moreover, the limited gap-opening decreases the chance to detect a forming planet.

\begin{table*}
  \caption{Mock continuum observations for various planet masses in ALMA Band 9 (440 microns) simulations}
 \label{tab:mock_sequence}
  \begin{tabular}{ccc}
  \hline
Saturn & 1$\mathrm{M_{Jup}}$   & 3$\mathrm{M_{Jup}}$\\
 \hline
   \includegraphics[width=0.7\columnwidth]{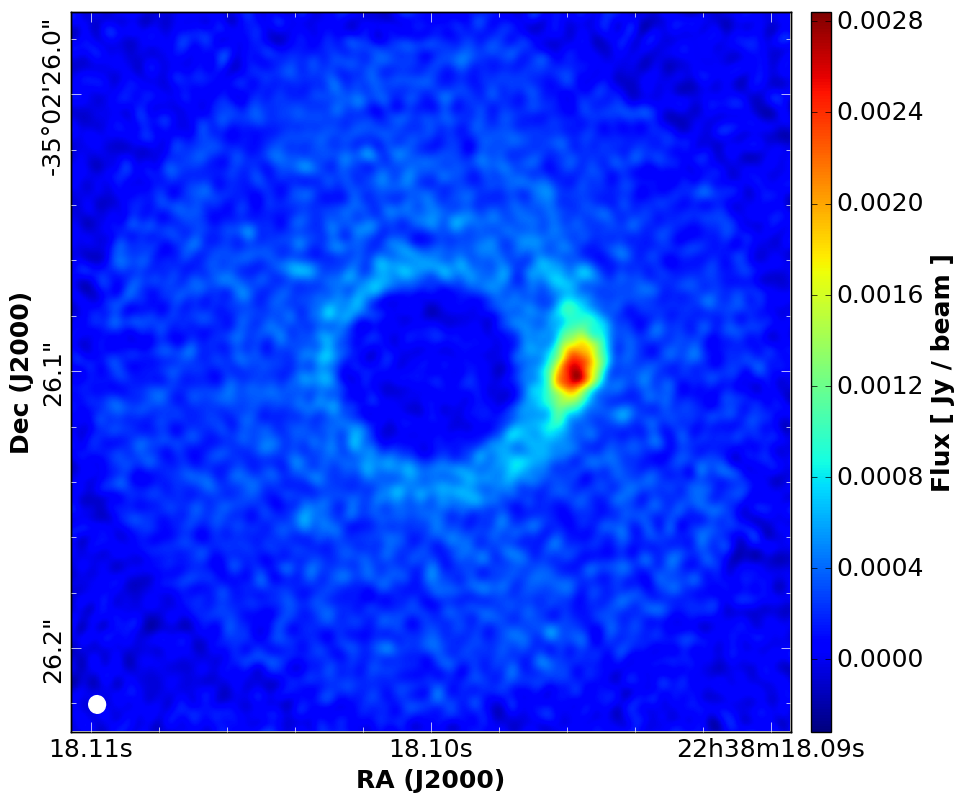}	&	\includegraphics[width=0.7\columnwidth]{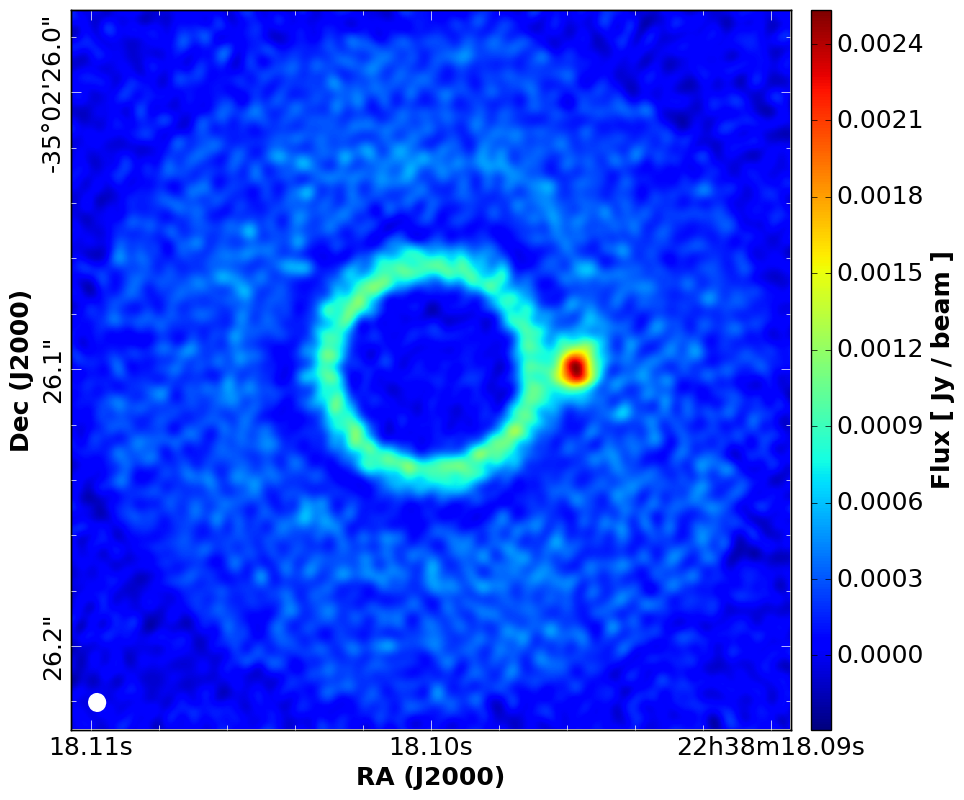}	&	\includegraphics[width=0.7\columnwidth]{res_440_3Jup.png}	\\
  \hline
 5$\mathrm{M_{Jup}}$ & 10$\mathrm{M_{Jup}}$ & \\
  \hline
\includegraphics[width=0.7\columnwidth]{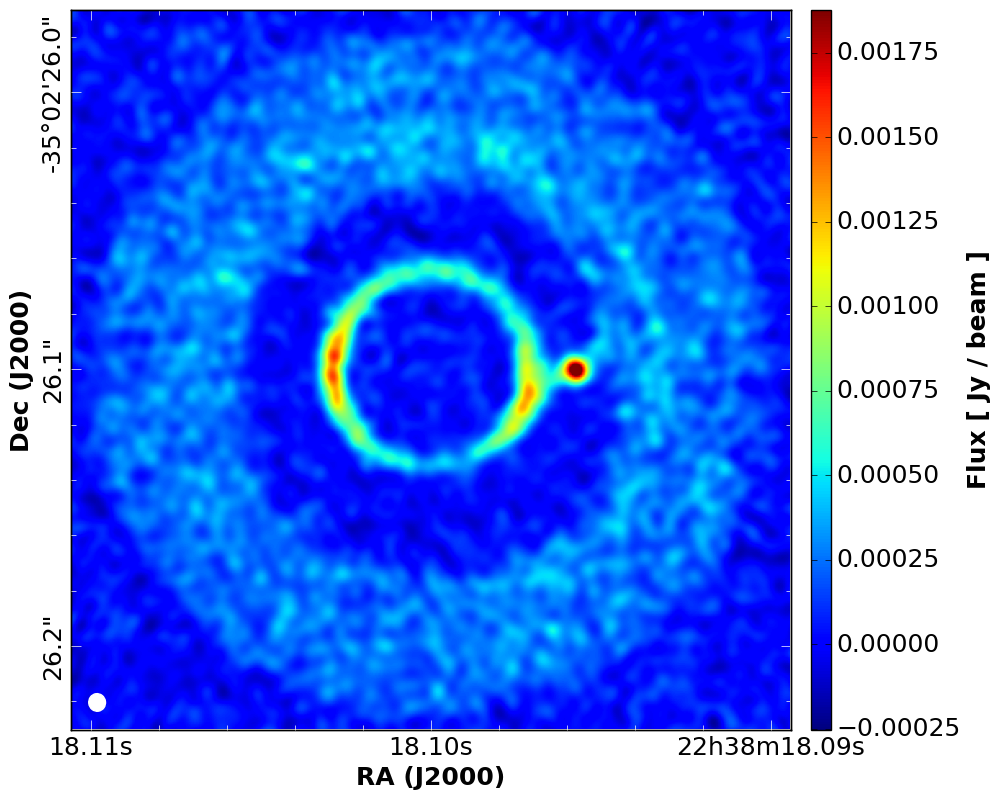} &\includegraphics[width=0.7\columnwidth]{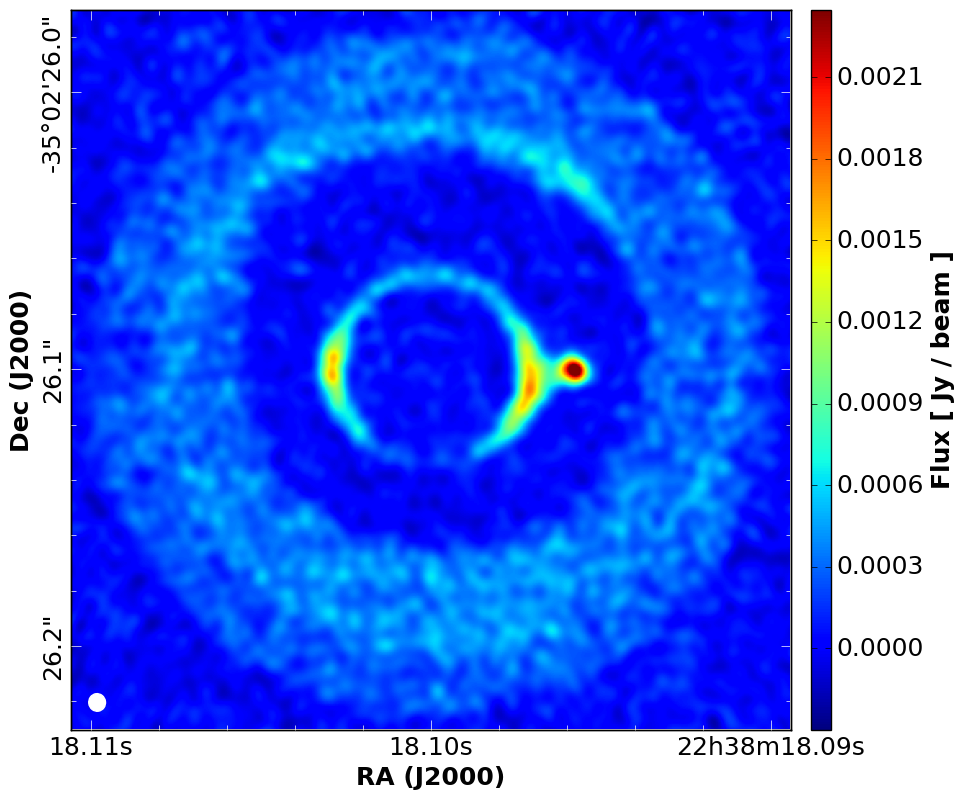} & \\
\end{tabular}
\end{table*}

For a successful detection, the contrast also matters between the (inner) circumstellar disk and the planet's vicinity. According to the figures in Table \ref{tab:mock_sequence} the inner protoplanetary disk seems fainter towards lower planetary masses. The reason for this is again the physics of gap-opening: giant planets push away the two parts of the circumstellar disk, i.e. pushing the inner circumstellar disk toward the star. Therefore the wider the planetary gap (the higher is the planetary mass), the higher the density in the inner circumstellar disk in the simulations (Table \ref{tab:mock_sequence}). Therefore, aiming for a better contrast ratio between the planet's vicinity and the inner circumstellar disk might be again better for lower planetary masses, although this can be influenced by several other mechanisms in a real disk, e.g. the accretion onto the star and the stellar magnetic field, just to name a few.

\subsection{The Role of Planet Temperature}
\label{sec:pltemp}

\begin{figure}
\includegraphics[width=\columnwidth]{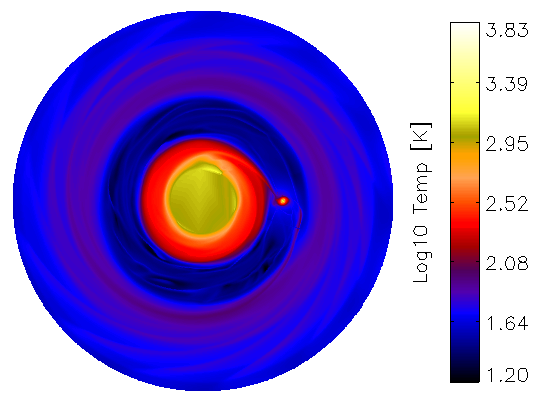}
\caption{Temperature map (natural base logarithmic scale) of the 3 Jupiter-mass hydrodynamic simulation. The planet vicinity pops out from the surroundings, due to the accretional- and viscous heating in the CPD and the planet irradiation. The inner protoplanetary disk is also hot due to the viscous heating and the higher optical depth (higher density). Mind the higher temperatures in the outer gap edge and in the spiral wake induced by the planet that are also due to the density/optical depth changes. These details, however, cannot be spotted in the ALMA synthetic images.}
\label{fig:temp_3jup}
\end{figure}

The temperature of the planet also influences the CPD temperature \citep{Szulagyi16a}, therefore its observability, as mentioned in Sect. \ref{sec:hydro}. In our simulations a planet temperature of 4000 K was used, so testing with a different value is important to understand how much a change in planet temperature is affecting the CPD observability. Therefore, a simulation was run with a 1 Jupiter-mass planet fixing the planet temperature to only 1000 K. In this case the ALMA CPD flux dropped by 15\% (band 10), 12\% (band 9), 19\% (band 8), 22\% (band 7), 27\% (band 6), and 29\% (band 4). This is roughly an increasing trend towards longer wavelengths, and hence the planet's irradiation is important for determining the CPD observability even in the sub-mm range, nevertheless the planet temperature is one of the least known parameters among the initial conditions. Most, currently available planet interior \& evolutionary models deliver effective temperature information only at 1 Myr or later, often neglecting the presence of a background disk. As it was shown in \citet{Szulagyi17}, the planet temperature influences the gas temperature even beyond the Hill-sphere, therefore to create realistic circumplanetary disk models, there is a need for planet interior models even within the first one million years of the formation. In any case, within this time period ($<$ 1Myr) the planet should have several thousand of Kelvin as a surface temperature, and given that the CPD directly touches this hot surface, it is inevitable that the sub-disk warms up significantly. Moreover, there is the accretional heating and the viscous heating which again heats the CPD. Despite the peak temperature of the CPD in our hydrodynamic simulations is nearly equal to that of the planet, which seems very high, but this is only true for the innermost few cells. The temperature gradient is very steep within the entire Hill-sphere (and even beyond), hence the temperature quickly drops to few hundreds of Kelvin in the bulk of the CPD in all our models (see Fig 5 and 6 in \citealt{Szulagyi17}). Nonetheless, the CPD is hotter in the optically thick limit than the majority of the circumstellar disk (except the inner rim of the protoplanetary disk) as it can be seen in Fig. \ref{fig:temp_3jup}.

\subsection{The Role of Orbital Separation}
\label{sec:orbital_sep}

The orbital separation of the planet can also significantly change the CPD observability. In our nominal simulations the planet is placed at Jupiter's location at 5.2 AU, but so far the detected embedded planetary candidates, e.g. in the LkCa15 system \citep{Sallum15,KI12}, the HD100546 planets \citep{Quanz15,Britain14}, the HD169142 candidates \citep{Reggiani14,Osorio14} are orbiting their stars much further away (between 15 and 52 AU). Therefore a simulation was run with the same setup as the nominal ones but placing the 1 Jupiter-mass planet at 52 AU, i.e., ten times further away. In this case the circumstellar disk that ranged between 21 and 125 AU. We found that the circumstellar disk and the CPD are both optically thin in all ALMA bands and in the hydro simulations Rosseland mean opacity assumption, that had a significant effect on the CPD temperature. The cooling time is almost instantaneous in comparison to the optically thick (closer-in planet) cases. This leads to that the bulk of the CPD is as cold as the surrounding circumstellar disk, only the very inner CPD ($<20$\% Hill-sphere, $<\frac{1}{3}\rm{R_{CPD}}$) is hotter than its surroundings. Nevertheless, this is still a significant gain in the surface area of the overheated region compared to the 5.2 AU planet simulations. The angular resolution requirement was tested in this 52 AU case -- i.e. whether the gap is resolved: the finding is that the resolution can be relaxed by a factor of ten (meaning 0.05 arcsec) in comparison to our nominal simulations. In fact, with the nominal 0.005 arcsec resolution the detection was not possible, as the synthetic image was purely noise.

In this large orbital separation case, higher sensitivity is needed for the CPD detection than in the nominal simulations. With 6 hours of \texttt{simobserve} integration time we were able to reach a clear, good detection (see Table \ref{tab:test_52au}). This means a theoretical noise limit (RMS) of  $5.3 \times 10^{-5}$ Jy in band 9, which according to the Sensitivity Calculator equals to five hours of integration time using the 15 GHz continuum bandwith. The theoretical RMS is $5.7 \times 10^{-6}$ Jy in band 7 and requires 9.5 hours of integration with 7.5 GHZ bandwidth. In band 9, the CPD peak flux is only 1.03 mJy/beam and the SNR relative to the gap and the disk is 6 and 4, respectively. Because the CPD is colder and optically thin in this case, band 7 gives almost the same good detection with $\rm{SNR_{gap}=5}$ and $\rm{SNR_{disk}=3}$ and peak flux of 0.30 mJy/beam. Note, however, that if the CPD is optically thick around 50 AU -- e.g. by having a more massive circumstellar disk -- the CPD observability should be significantly better in band 9 simply because of the hotter CPD due to the longer cooling timescale.

\begin{table*}
  \caption{Test for large orbital separation and optically thin CPD -- Jupiter-mass planet at 52 AU}
 \label{tab:test_52au}
  \begin{tabular}{cc}
  \hline
Band 9 & Band 7 \\
$\rm{RMS_{theoretical}} = 5.3 \times 10^{-5}$ Jy & $\rm{RMS_{theoretical}} = 5.7 \times 10^{-6}$ Jy\\
$\rm{RMS_{measured}} = 2.7 \times 10^{-4}$ Jy &  $\rm{RMS_{measured}} =  1.5 \times 10^{-5}$ Jy \\
 \hline
   \includegraphics[width=0.9\columnwidth]{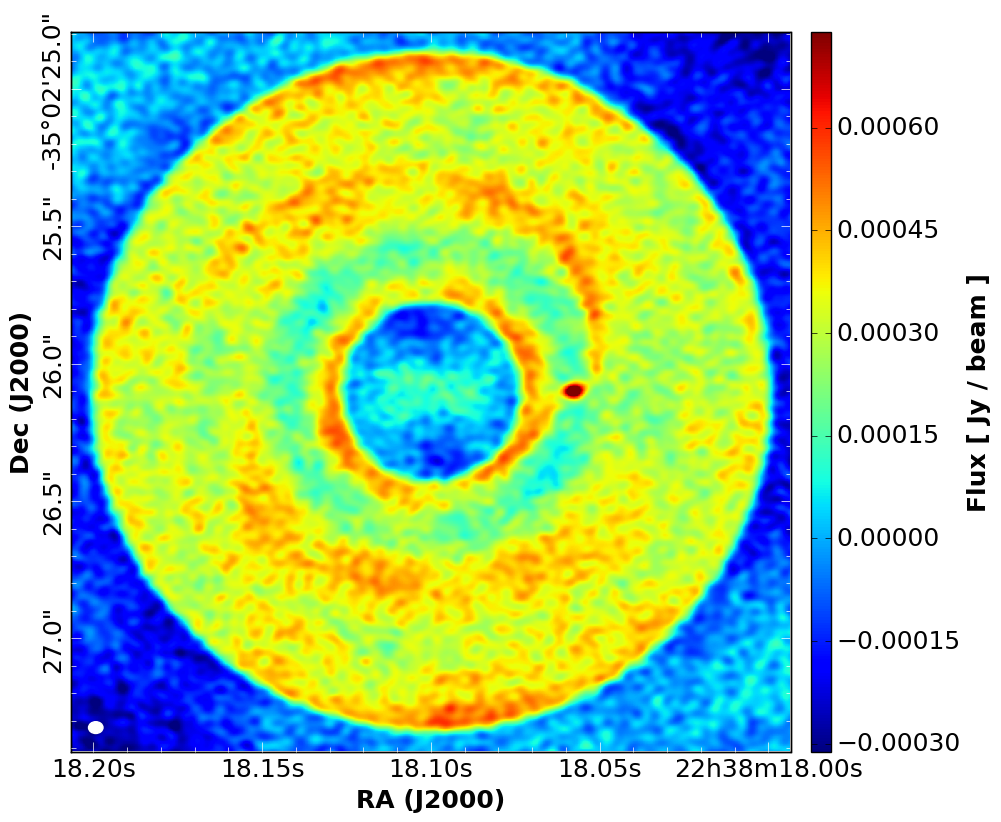}	&	\includegraphics[width=0.9\columnwidth]{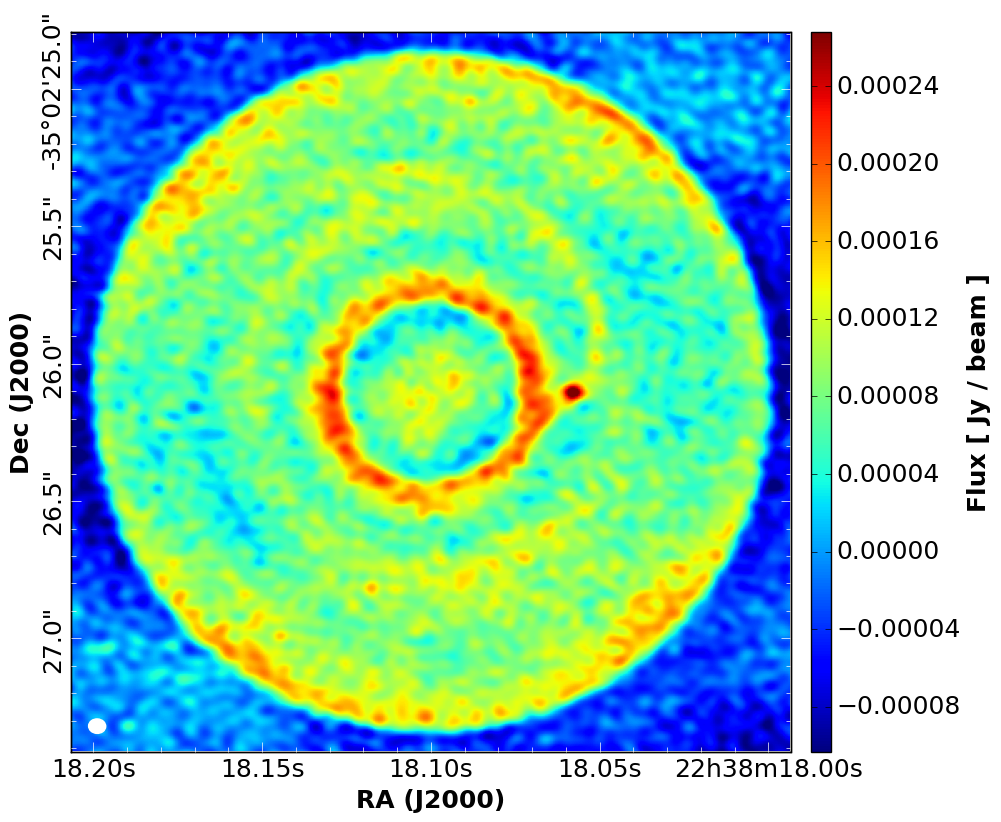}		\\
  \hline
\end{tabular}
\end{table*}

\subsection{Gap Profiles}

\begin{figure}
\includegraphics[width=\columnwidth]{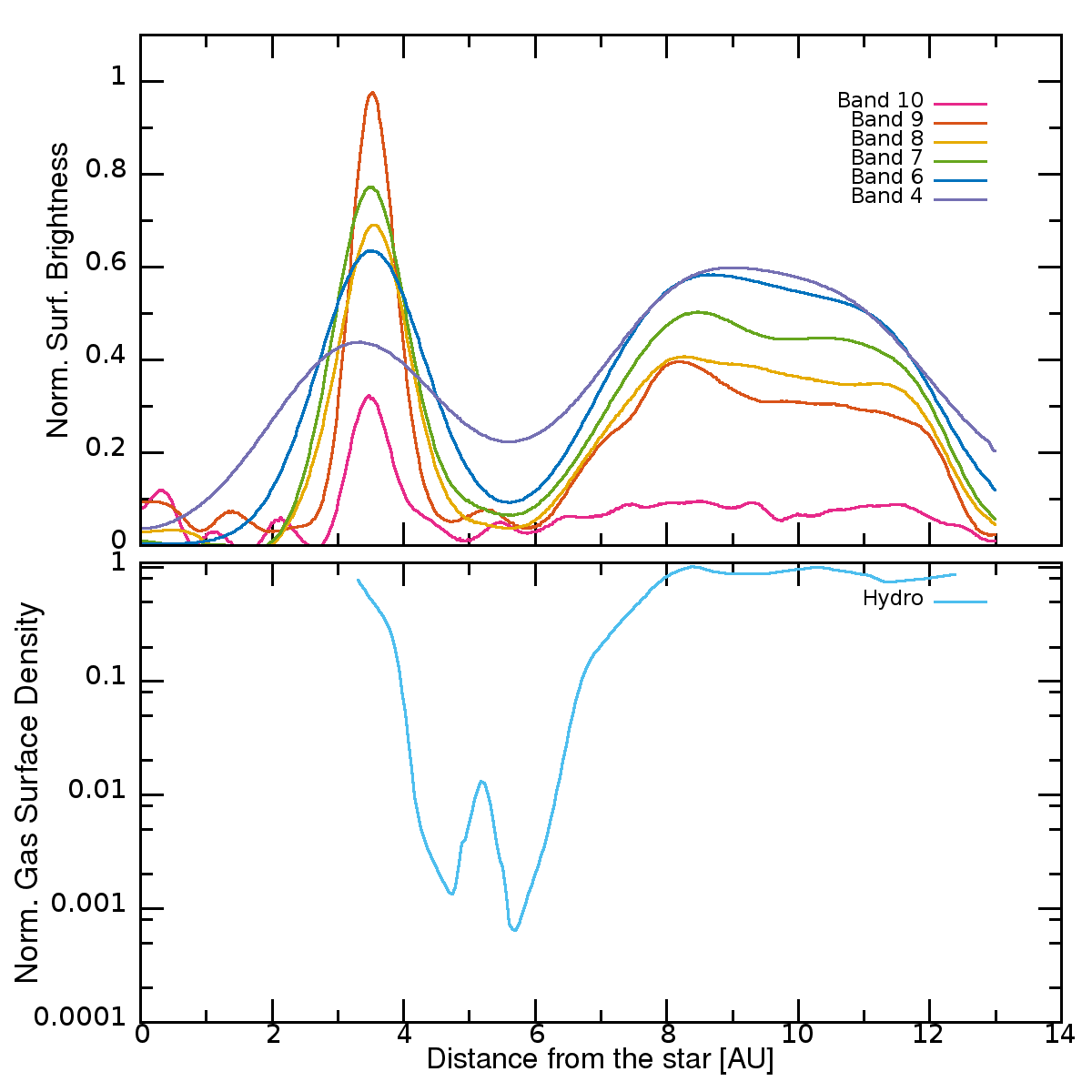}
\caption{Azimuthally averaged gap profiles for ALMA bands 4-10 (top) in the simulation with the 5 Jupiter-mass planet and the hydro simulation azimuthally averaged profile (bottom). Clearly, the gap width and depth is quite different at the various ALMA bands and significantly different in the hydrodynamic simulation. Therefore, one should pay extra attention when using ALMA observations of gaps to estimate planetary masses.}
\label{fig:gaps}
\end{figure}

Given that our radiative hydrodynamic simulations are global simulations, i.e. they contain a big ring of the circumstellar disk around the gas giants, the planetary gaps can also be examined. The observed gap widths and depths are often used directly to estimate the planetary mass, by using analytical formulae \citep{Crida07,Duffel15} that were created from non-radiative gas hydrodynamic simulations. However, the gap-profile is significantly different for the same  planet in different bands, partially because of the different optical depths (gap is wider and deeper with longer wavelengths), and partially because of beam convolution (beam size is larger at longer wavelengths), as it is shown in Fig. \ref{fig:gaps}. We azimuthally averaged the surface brightness of the mock observations of the 5 Jupiter-mass simulation in the various ALMA bands (top panel of the figure), normalizing with the maximal value 0.827 mJy/beam). For comparison, the the gap profile of the hydrodynamic simulation is also plotted in the bottom panel with normalizing the curve with the maximal surface density. The gap depth is 3 orders of magnitude in the hydro simulation, but a factor of a few in the ALMA mock observations. The gap width, which is the more reliable parameter to estimate the planetary mass, is $\sim$ 2 AU wide in the hydro simulation compared to 1-1.5 AU in the mock observations. This 25-50\% difference in the gap width can significantly underestimate the planet mass that carved the gap \citep{Crida07,Duffel15}. In conclusions, the advise is to couple dust included 3D hydrodynamical models with wavelength dependent radiative transfer and the ALMA simulator to estimate the planet masses from ALMA observed gap profiles.

\section{Discussion}

It has been studied since a long time that dust gaps opened by embedded planets differ in width and depth from gas gaps \citep[e.g.][]{Paardekooper04,Paardekooper06,Rosotti16,Ruge16,Zhu14,Fouchet10}. While only giant planets ($\sim$ Saturn or more massive) can open gas gaps, even lower mass planets can open dust gaps. In the cases where a gas gap is opened, the corresponding dust gaps (depending on the grain size, and, whether the dust is decoupled from the gas) should be wider and deeper. This scales with the grain size: the larger are the dust particles, the wider the dust gap will be. In our simulations, only gas is included but we were making predictions for the observed dust distribution, therefore it is crucial to test whether a strong dust and gas coupling can be assumed. ALMA dust continuum images are sensitive to $\sim$ millimeter sized grains, therefore the Stokes number was calculated for such particle sizes. As it is discussed in Sect. \ref{sec:radmc3d}, the Stokes number $\ll$ 1.0, therefore strong coupling can be assumed. However, some caution has to be made when interpreting the results without proper dust treatment in the hydrodynamic simulations.

In this work a fixed dust-to-gas ratio of 0.01 was used, however, this value should change within the circumstellar disk, for example at dust traps \citep[e.g.][]{YG05,Marel13,DD14,Birnstiel12,Meheut12}. Furthermore, recent surveys with ALMA revealed that the dust-to-gas ratio varies between 0.1-0.001 in various circumstellar disks \citep[e.g.][]{WB14,Ansdell16}. Due to the fact that the CPD is fed mainly from the vertical influx of the meridional circulation, only the small sized dust, that is well coupled to the gas can enter the CPD. Centimeter and larger sizes are stopped at the pressure bump of the outer gap edge \citep{Pinilla2015b}. This means that the dust coagulation has to start over inside the CPD, which could result in a significantly different dust-to-gas ratio compared to that of the circumstellar disk.

The forming planet temperature matters a lot for observability, as it was showed in Sect. \ref{sec:pltemp}. Planet interior and evolution studies are mostly made for fully fledged planets, i.e. after they detach from the surrounding disk. However, the planet temperature should be known for realistic CPD studies, because unlike accretion onto stars, the disk of planets are touching the planetary surface. Therefore, the heating of the planet is essential for the thermodynamics of the CPD. This emphasizes that planet interior and evolution models are needed for the formation phase too.

The fact that there is no unambiguous CPD detection is surprising, because every second star should have a giant planet between 5-20 AU \citep{Bryan16}, and every forming gas giant should be surrounded by a CPD. The question is why there is no CPD detection so far if there is a vast amount of these disks out there. First, they are only 0.3-0.5 Hill-radii \citep[e.g.][]{Szulagyi14,Szulagyi16a,Tanigawa12}, therefore resolving them is not possible today. Due to the ongoing accretion from the circumstellar disk onto the CPD and the planet, distinguishing between the planet and its disk will be challenging. Third, even though a couple of forming planet candidates were found with direct imaging technique, the confirmation that they are indeed planets with disks around them, instead of some other disk feature is under progress. If these planetary candidates are in fact not planets, it is understandable why their CPD was not detected so far. Furthermore, low mass planets and hot gas-giants \citep{Szulagyi16a} will not form disks, only envelopes around them. Their dynamical imprint therefore should be significantly different than of a disk (e.g. significantly smaller rotation in comparison to a disk). In conclusion, detecting the CPD still remain a challenging task, even if they should be frequent.

\section{Conclusions}

In this paper mock observations for ALMA continuum data are presented that were created from 3D radiative hydrodynamic simulations of forming planets embedded in circumstellar disks. Our main goal was to investigate whether the circumplanetary disk (CPD) formed around the planet can be detected with ALMA, and if yes, which band is the best, and what are the angular resolution and sensitivity requirements. 

Simulations of Saturn, 1, 3, 5 and 10 $\mathrm{M_{Jup}}$ planets were carried out. All these gas giants orbit a Solar-mass star at 5.2 AU, but tests were also run for an orbital separation of 52 AU and two planet temperatures (4000 K and 1000 K).

Our parameter study revealed that the CPD is hotter than most of the circumstellar disk, due to accretional heating, planet irradiation, shock- and viscous heating. When the planet is at 5.2 AU, the CPD is optically thick, therefore the cooling time is very long. Hence, shorter wavelengths are better to target CPD observations in the continuum, preferably Band 9 and 10. At 440 microns (band 9) only one hour of integration time is needed for a detection with an SNR of (minimum) 4. The resolution and the contrast ratio with the hot inner rim of the protoplanetary disk is also the most favorable in these bands. It was also found that, surprisingly, the band 7, 8, 9 fluxes in the planet vicinity in the case of lower mass planets (Saturn and one Jupiter-mass gas giants) are about the same or higher than in the case of a 10 Jupiter-mass planet. This is due differences in temperature weighted optical depths: the larger the planetary mass, the wider and deeper the gap, therefore the planet's vicinity is more optically thin, the cooling of the CPD is more efficient. On the other hand, in the case of the Saturn-, and Jupiter-mass planets, the gap is shallower and narrower, there is more gas and dust around the planet even beyond the Hill-sphere, therefore a larger area around the planet is heated up due to the accretion process and the planet's irradiation. In these lower mass cases, the contrast relative to the inner circumstellar disk is also better, because the planets do not open that wide gap, that would push more mass into the inner protoplanetary disk causing it to be brighter in the ALMA bands. However, this all means that detecting hot dust around planets does not necessarily mean a CPD detection. In the case of imperfect (or no) gap-opening, simply the circumstellar disk dust heats up around the planet, leaving a large area ($>\rm{R_{Hill}}$) overheated around the small mass planet. 

Among the gas giants that carve deep gaps (3, 5, 10 Jupiter-mass planets in this study), there is a correlation between the ALMA flux and the planetary mass: e.g. the vicinity of the 10 Jupiter-mass planet is brighter than that of the 3 $\mathrm{M_{Jup}}$, due to the higher peak densities reached near the planet. It was also showed that the ALMA flux will depend on the assumed planet temperature, that is unknown for forming planets within the first million years; therefore there is a need for such planetary interior \& evolution models.

The orbital separation of the planet also significantly changes the observability. A test simulation of a Jupiter-mass planet at 52 AU was also perfored, leading to an optically thin CPD. Due to this fact, the cooling is much more efficient and only the inner one-third of the CPD is overheated with respect to the surrounding circumstellar disk. As a result, in this large orbital separation case, a significantly higher sensitivity was needed (5 hrs integration time) to detect the CPD around this planet than if the gas-giant was at 5.2 AU. The resolution can be relaxed by a factor of 10, and due to the colder CPD the contrast ratio with the circumstellar disk is similar in band 7 as in band 9.

Comparing the gap profiles of the same simulation (same planet mass) in the different ALMA bands confirmed that the gap depth and width drastically changes with the wavelength for mainly two reasons: the gap is wider at longer wavelengths due to decreasing optical depth, and, the convolution with the different beam sizes will increase the gap depth. It was showed that the hydrodynamic simulation gap is 25-50\% wider than in the ALMA mock observations. Therefore, extra caution has to be made when estimating planet masses based on ALMA continuum observations of gaps.

\section*{Acknowledgments}

We thank for the insightful review for the anonymous referee and for the useful discussions to A. Juh\'asz and. J. Sz. acknowledges the support from the ETH Post-doctoral Fellowship from the Swiss Federal Institute of Technology (ETH Z\"urich). This work has been in part carried out within the frame of the National Centre for Competence in Research  ``PlanetS"  supported by  the  Swiss  National Science Foundation. S.D. and S.P.Q. acknowledges the financial support of the SNSF. Computations have been done on the ``M\"onch" machine hosted at the Swiss National Computational Centre.

\label{lastpage}

\end{document}